\documentstyle[12pt]{article}

\newcommand{\be}{\begin{equation}}
\newcommand{\ee}{\end{equation}}
\newcommand{\bea}{\begin{eqnarray}}
\newcommand{\eea}{\end{eqnarray}}
\newcommand{\beann}{\begin{eqnarray*}}
\newcommand{\eeann}{\end{eqnarray*}}
\newcommand{\beasn}{\begin{sneqnarray}}
\newcommand{\eeasn}{\end{sneqnarray}}

\newcommand{\eps}{\epsilon}

\newcommand{\tg}{\Delta}
\newcommand{\nn}{\nonumber}
\newcommand{\der}[2]{\frac{\partial #1}{\partial #2}}

\newcommand{\lra}{\leftrightarrow}
\catcode`@=11
\def\dif{{\rm d}}
\def\deriv{\@ifnextchar[{\@deriv}{\@deriv[]}}
   \def\@deriv[#1]#2#3{\mathchoice%
{{\dif^{#1}#2\over\dif{#3}^{#1}}}{{\dif^{#1}#2/\dif{#3}^{#1}}}%
{{\dif^{#1}#2\over\dif{#3}^{#1}}}{{\dif^{#1}#2/\dif{#3}^{#1}}}}

\def\presup#1{{}^{#1}\kern-.15em\relax}      %pre-superscript
\def\presub#1{{}_{#1}\kern-.12em\relax}      %pre-subscript

\def\relstack#1#2{\mathrel{\mathop{#1}\limits_{#2}}}
\def\secteqno{\@addtoreset{equation}{section}%
\def\theequation{\thesection.\arabic{equation}}}
\def\endsecteqno{\def\theequation{\@ifundefined{chapter}%
{\arabic{equation}}{\thechapter.\arabic{equation}}}}
\newcounter{subequation}
\def\thesubequation{\alph{subequation}}
\def\sneqnarray{\stepcounter{equation}\let\@currentlabel=\theequation
\setcounter{subequation}{1}
\def\@eqnnum{{\rm (\theequation\thesubequation)}}
\global\@eqcnt\z@\tabskip\@centering\let\\=\@eqncr\let\@@eqncr=\@@sneqncr
$$\halign to \displaywidth\bgroup\@eqnsel\hskip\@centering
 $\displaystyle\tabskip\z@{##}$&\global\@eqcnt\@ne
 \hskip 2\arraycolsep \hfil${##}$\hfil
 &\global\@eqcnt\tw@ \hskip 2\arraycolsep $\displaystyle\tabskip\z@{##}$\hfil
  \tabskip\@centering&\llap{##}\tabskip\z@\cr}
\def\endsneqnarray{\@@sneqncr\egroup $$\global\@ignoretrue}
\def\@@sneqncr{\let\@tempa\relax
   \ifcase\@eqcnt \def\@tempa{& & &}\or \def\@tempa{& &}
   \else \def\@tempa{&}\fi
     \@tempa \if@eqnsw\@eqnnum\stepcounter{subequation}\fi
     \global\@eqnswtrue\global\@eqcnt\z@\cr}
\def\nobiblabels{\def\@lbibitem[##1]##2{\@bibitem{##2}}}
\catcode`@=12
\secteqno

\title{\bf\Large   Regularization Methods in Chiral Perturbation Theory\\}

\author{\sc  D.Espriu\thanks{E-mail address: espriu@ebubecm1.bitnet} and
J. Matias\thanks{E-mail address: matias@ebubecm1.bitnet}\\
	{\normalsize D.E.C.M.} \\ {\normalsize Universitat de Barcelona} \\
	{\normalsize Diagonal 647, 08028 Barcelona, Catalonia, Spain}}
\begin{document}
\date{}
\maketitle
\thispagestyle{empty}
\begin{abstract}

Chiral lagrangians describing the interactions of Goldstone bosons
in a theory possessing spontaneous symmetry breaking are
effective, non-renormalizable field theories in four dimensions.
Yet, in a momentum expansion one is able to extract definite, testable
predictions from perturbation theory. These techniques
have yielded in recent years a
wealth of information on many problems where the physics of
Goldstone bosons plays a crucial role, but theoretical issues
concerning chiral perturbation theory remain, to this date, poorly
treated in the literature.
We present here a rather comprehensive analysis of the regularization and
renormalization ambiguities appearing in chiral perturbation theory
at the one loop level. We discuss first on the relevance of dealing with
tadpoles properly. We
demonstrate that  Ward identities severely constrain the choice of regulators
to the point of enforcing unique, unambiguous results in chiral
perturbation theory at the one-loop level for
any observable which is renormalization-group invariant.
We comment on the physical implications of these results and on
several possible regulating methods that may be of use for some
applications.
\end{abstract}

\vspace{-195mm}
\vfill\hfill
\vbox{
\hfill UB-ECM-PF 93/15\null\par
\hfill June 1993}\null
\vspace{195mm}

\clearpage

\section{Introduction}

The phenomenon of the spontaneous breakdown of a continuous
symmetry is one of the recurrent themes in modern physics.
Randomly chosen examples exhibiting this property are
the dynamical breakdown of the chiral symmetry in Quantum
Chromodynamics\cite{GaLe1,GaLe2}, the BCS
theory of superconductivity\cite{WeRe},
the breaking of the electroweak symmetry that gives masses to
the $W^\pm$ and $Z$ intermediate vector bosons\cite{WeOl}, or the
appeareance of the N\'eel state in strongly coupled electron
systems\cite{HiTc}.

A characteristic signal of the spontaneous breakdown of a
continuous symmetry
is the appeareance  of Goldstone bosons\cite{Gold}.
This is a very general
phenomenon, taking place
both for relativistic and non-relativistic systems. Stated in
mathematical terms, it means that there are states whose energy
vanishes as the three-momentum tends to zero.
The dispersion relation connecting energy and momentum can be either
linear
($E\sim \vert k\vert$, for a relativistic theory), or
non-linear
($E\sim k^2$, for non-relativistic systems).
 From the point of view of field theory this amounts to stating the
presence of
zero-mass particles in the spectrum.

The implications of the spontaneous breakdown of a continuous symmetry
go, however, well beyond the mere appeareance of Goldstone bosons. The
interactions of Goldstone bosons are largely dictated by very
general considerations. On the one hand, interactions of Goldstone
bosons necessarily contain at least one derivative. This can be
understood as follows. The  effective Hamiltonian describing the Goldstone
bosons must be invariant under constant shifts of the Golstone fields
\be
\pi(x) \to \pi(x)+ c
\ee
since such a field redefinition does not change the energy-momentum
dispersion relation. The above invariance is ensured by derivative
couplings. The fact that interactions are proportional to momenta
implies that at zero momentum Goldstone bosons do not interact.

On the other hand, symmetries, even if they are partially
broken, are powerful enough to dictate the form of the effective
long distance lagrangian for these Goldstone modes. Since the
work of Callan, Coleman, Wess and Zumino \cite{CaCo} we know that an economic
and handy way of describing the interactions of Goldstone bosons
is provided by a non-linear sigma model. The set of Goldstone bosons
appearing after the breaking of a symmetry group $G$ to some subgroup $H$
can be grouped in a dimensionless
matrix-valued field $U(x)$ belonging to the quotient space $G/H$. The
effective lagrangian can be organized as a momentum expansion
with higher dimensional operators containing an increasing number of
derivatives acting on the field $U(x)$.

The lagrangian governing the physics of Goldstone bosons is, except in very
simple cases, non-linear.  Thus we must deal
with theories which are not very tractable from a quantum
point of view in four dimensions. Can we, perhaps, ignore quantum
corrections? The answer is
except if we content ourselves with the crudest of approximations, no.
The tree level approximation will be good
only at very
low energies. Loops give sizable contributions to many observables
 and, characteristically,
they are amongst the most interesting ones.
Perhaps
the most striking examples are those where a process is actually
forbidden at tree level, but allowed by loop corrections, such as
the rare (but measured) decay
$K_S\to \gamma\gamma$ \cite{EsGi} or, in
the process $\gamma\gamma\to Z_L Z_L$, recently proposed \cite{Expe} as
an excellent way
of testing possible departures from the minimal Standard Model\cite{MariJo}.
(Using the so-called
equivalence theorem\cite{EqTh}, the scattering of longitudinal vector bosons
can be expressed in terms of an equivalent process with the Goldstone
bosons of the broken electroweak symmetry.)

 We shall be concerned here with a particular pattern of
 symmetry breaking, in which the symmetry group  $G=SU(N)_L\times
 SU(N)_R$ is broken to its diagonal subgroup $H=SU(N)_V$. Originally,
 we are dealing here with a lagrangian that depends
 on some matrix-valued field $M(x)$, which is left
 invariant under independent
 left and right multiplication of $M(x)$ by constant
matrices of $SU(N)$. When this symmetry is broken to the
diagonal group $SU(N)_V$, there appear $N^2-1$ Goldstone
bosons,  grouped in a matrix-valued field $U(x)$
belonging to the group $SU(N)_L\times SU(N)_R/SU(N)_V$.
This breaking pattern, setting $N=3$, is the one relevant
in the realization of chiral symmetry in Quantum Chromodynamics at low
energies.
$N=2$ corresponds to
the breaking pattern of the global symmetry of the scalar sector of
the Electroweak Theory, responsible for the generation of mass of the
$W^\pm$ and $Z$ particles.
The longitudinal
components of these massive particles are precisely
the Goldstone bosons associated to the broken generators.
The effective lagrangians that correspond to the above
breaking pattern are usually named chiral lagrangians.

The number of operators in the effective chiral lagrangian
with the right invariance properties is relatively small if we do not
consider the coupling to external fields and we assume that
there are no terms that break explicitly the original chiral
group $SU(N)_L\times SU(N)_R$. There
is only one operator with two derivatives
\be
\label{L2}
{\cal L}^{(2)}=\frac{f_\pi^2}{4} Tr \partial_\mu U^\dagger
\partial^\mu U
\ee
There are three operators with four derivatives\cite{GaLe2}
\bea
\label{L4}
&{\cal L}^{(4)}=L_1 Tr(\partial^{\mu} U^{\dag} \partial_{\mu}
U)^2+L_2
Tr(\partial_{\mu} U^ {\dag} \partial_{\nu} U) Tr(\partial^{\mu}
U^{\dag} \partial^{\nu}
U)&\nn\\ &+L_3 Tr(\partial^{\mu} U^{\dag} \partial_{\mu} U
\partial^{\nu} U^{\dag} \partial_{\nu} U)& \eea
and
operators proliferate if we consider
terms with six or more derivatives.

Notice that there is a dimensional constant $f_\pi$ in front of
(\ref{L2}). In pion physics, $f_\pi$ is the pion
decay constant. The very presence of this dimensional
coupling constant plays a crucial role in
chiral
perturbation theory. Since we have a dimensional parameter at our disposal
we can construct, with the help of the momenta,
a {\it dimensionless} quantity $q^2/(4\pi f_\pi)^2$
which can be made arbitrarily small. We therefore have an adjustable
parameter allowing us to determine a range of energies in which
tree level is certainly more important than one-loop corrections, and
these in turn are more important than two-loop corrections and so on.
(In two dimensions, on the contrary,  the equivalent to $f_\pi$ is
dimensionless and we are left without any adjustable parameter.
It is thus remarkable that two-dimensional chiral models are actually
much more difficult to treat than four dimensional ones.)

$f_\pi$  sets an absolute scale with respect to which
one measures all energy scales. Any effective
theory possessing the same long-distance properties will necessarily lead to
the same operator ${\cal L}^{(2)}$, with the same value for the coefficient
in front. Because
${\cal L}^{(2)}$ is completely general, if we are somehow limited to
performing
low energy experiments we will learn
very little, if anything at all, about the underlying theory from measuring
$f_\pi$ very precisely. This is very clearly seen in attempts to derive
a long distance effective lagrangian from QCD. This is, needless to say,
a very difficult problem; a strongly coupled theory deep in the
non-perturbative regime. Short of solving QCD exactly we can try to make
some
more or less justified simplifications, in the hope that we are somehow
modelling
QCD in an intermediate energy range. For instance, we can assume that,
for some range of energies, interactions
are dominated
by the interchange of a scalar particle. Or, we can instead assume
that vector and axial-vector particle interchange does dominate
long distance interactions. The point is that in both cases we can
always tune some parameters so as to reproduce the experimental
value of $f_\pi$. How can we tell which of the two
models  is more correct? Certainly
measuring  $f_\pi$ will be of no help in this
issue.

On the contrary,
it has been found that integrating out the
scalar, or the vector and axial-vector effective degrees of
freedom lead to very different values for the ${\cal O}(p^4)$
coefficients. The values obtained from vector and axial-vector
interchange are, on the whole and roughly speaking, much more compatible
with the
experimental results than those obtained from scalar interchange
alone\cite{PiRe}. So there
is definitely some truth in assuming that at some intermediate scale
massive vector particles dominate strong interactions.
In the old days of hadronic physics this very fact used to be
called vector-meson dominance\cite{VeMe}.

We would certainly like to find the values of these
constants $L_i$ directly from the QCD lagrangian or, generally
speaking, for any theory that exhibits chiral symmetry
breaking because, as we have just discussed, they contain the relevant
information at low energies about
whatever underlying physics is behind. For instance, it would be of utmost
interest to be able to compare the theoretical and experimental values for
these coefficients in the symmetry breaking sector of the Standard
Model to verify or falsify different models for the scalar sector. A lot
of work has been done recently in this
field\cite{Longhi,Appel,EsHe,Peskin,Holdom}. But {\it en rigueur} there are
several theoretical issues that need be carefully clarified before one is
entitled to make a detailed comparison with experiment and jump into
conclusions.

One of the reasons is that
the chiral lagrangian
is a non linear sigma model and
it has a very
bad ultraviolet behaviour in perturbation theory. Naive power counting
allows even for quartic divergences and quadratic
as well as logarithmic divergences are present\cite{Georgi,WePhy}. As often
happens in field theory, symmetries must play a crucial role in determining,
how many of the possible counterterms that are allowed by naive  power
counting do actually appear. Even after taking this into
account it should be obvious that quantum corrections in a non-linear
theory of this type are badly divergent.
In fact non-linear
sigma models in four dimensions are not renormalizable in perturbation
theory. This means that if we start with the simplest non-linear
theory, the one described by ${\cal L}^{(2)}$, and we compute loops
with it, new counterterms will be required at each
order in perturbation theory. In particular, it will certainly
be necessary to redefine the coefficients $L_i$ of ${\cal L}^{(4)}$
order by order in chiral perturbation theory to absorb some logarithmic
ultraviolet divergences. In principle this is not very different from
what one does in renormalizable theories, except that the number of
counterterms here is, strictly speaking, infinite. However, at
any giver order in the adjustable parameter $q^2/(4\pi f_\pi)^2$
the number of possible counterterms is finite.

The renormalized parameters can be obtained
 from experiment. From comparison with the experimental data one
will learn that, for instance, one of the $L_i$'s
takes a given value when working in a given
regularization and at a given subtraction scale $\mu$. Let's call the
renormalized coefficients $L_{i}(\mu)$.
The double dependence both on the regulator and on the scale means that
none of the $L_i(\mu)$'s have {\it per se} any physical meaning. Of
course,
for practical applications this is just fine. If we work consistently in
the same regularization and renormalization scheme we can use the value
of $L_i(\mu)$ just obtained in other processes
and make definite, testable predictions.
This is the standard procedure in pion physics\cite{GaLe1}.

However, rather than just fitting the coefficients of the
chiral lagrangian
we would like to {\it compute} them. We would like to determine
the values of, say, one of the $L_i$ in a given
theory and compare this values with the data available.
In order to do that we need a good theoretical control of the
precise relation between those values of the renormalized coefficients
$L_i(\mu)$
that can be extracted from the experiment and those that we can compute
in a particular model. That the issue is far from obvious can be
illustrated
 from recent estimations on the values of the ${\cal O}(p^4)$
coefficients
 from QCD. After integrating out the quark and gluonic degrees of
freedom, at leading order in $1/N_c$, $N_c$ being the number of colors,
the authors of \cite{ERT} found the following estimation for $L_1$,
$L_2$ and $L_3$
\be
\label{estim}
L_1=\frac{N_c}{384 \pi^2}\qquad L_2=\frac{N_c}{192\pi^2}\qquad
L_3=-\frac{N_c}{96\pi^2}
\ee
The leading gluonic corrections turn out to be zero for $L_1$ and
$L_2$, but non-zero for $L_3$. Let's ignore the gluonic corrections
altogether for the present discussion. The point we need to retain is
that the above values for the $L_i$ are finite, non-ambiguous and of
${\cal O}(N_c)$ and remain so after including ${\cal O}(N_c)$ gluonic
corrections. Furthermore they appear to be independent of the
regularization scheme used to derive them \cite{ERT}. Now we want to
compare
these theoretical predictions with the experimental values for
$L_1(\mu)$, $L_2(\mu)$ and $L_3(\mu)$. The
latter are, however, renormalized coefficients that depend on some
subtraction scale and on the regulator that has been used to render
quantities finite in chiral perturbation theory. Is the comparison
meaningful, or even possible?

 From large $N_c$ counting arguments it is not difficult to show that
$f_\pi^2$ is of ${\cal O}(N_c)$; therefore contributions from chiral
loops (including the scale and regulator ambiguities)
are down by powers of $1/N_c$. It is also known from chiral
perturbation theory that
there are combinations of renormalized coefficients, such as
$L_1(\mu)-\frac{1}{2}L_2(\mu)$ which
are renormalization group invariant at the one-loop level from the
point of view of chiral
lagrangians, i.e. where the logarithmic divergences of the effective
theory cancel
\be
\label{combi}
\mu {\partial \over {\partial\mu} } (L_1(\mu)-{1\over 2}L_2(\mu)) =0
\ee
The combination $L_1(\mu)-\frac{1}{2}L_2(\mu)$ is not an observable by
itself and in principle it needs not be regulator independent.
In physical amplitudes is always
accompannied by a finite piece generated by one-loop perturbation
theory; i.e. by intermediate states with at least two Goldstone bosons.
Let us denote by $F$ this
finite piece, which is  of ${\cal O}(1)$ in the $N_c$ expansion. The
combination $A=F+L_1(\mu)-\frac{1}{2}L_2(\mu)$
is observable and, hence, should be independent of the
regulator.
In the best of worlds we could determine this observable quantity F from
first principles, i.e. from QCD.
Then, after fixing
the only free scale in the theory, $\Lambda_{QCD}$, from
$f_\pi$, and expanding the amplitude up to ${\cal O}(p^4)$ we would
find a unique, non ambiguous, value for the observable  $A$. This
value could then be expanded in powers of $1/N_c$. From large
$N_c$ arguments we know that the leading contribution to the
effective
action is of order ${\cal O}(N_c)$, corresponding to contributions from
intermediate states with two quark lines. At order $p^4$ it would be given
by the coefficients
$L_i$ in eq. (\ref{estim}). In addition, although they have never
been explicitly computed, there are contributions
 from intermediate states with four or more quark lines .  These are of
${\cal
O}(1)$ and must correspond, in the effective chiral theory, to $F$.
Therefore, while we cannot, strictly speaking, compare the values for
$L_1(\mu)$
or $L_2(\mu)$ that we get from experiment with the theoretical
predictions
of \cite{ERT} because the former have arbitrary subtractions that depend on
the calculational scheme we use in chiral perturbation theory and have
nothing to do with the fundamental theory, we expect to
be able to compare the theoretical
and experimental predictions for the renormalization-group
invariant combination $L_1(\mu)-\frac{1}{2}L_2(\mu)$.
We should mention that from \cite{ERT} one concludes that, up to
subleading gluonic corrections \be
L_1-{1\over 2}L_2 \simeq 0
\ee
In fact, it is known that in $\pi\pi$ scattering the $J=1$, $I=1$
amplitude is very well described if one
ignores altogether the contributions from the ${\cal O}(p^4)$
operators, which precisely turn out to be proportional to the
combination $L_1-\frac{1}{2} L_2$.

For the above identification to work a necessary condition must be met.
The value for the finite contribution $F$ that comes from the one loop
chiral expansion must be regulator independent. It must be a finite,
unambiguous number.
Otherwise it would be very difficult to attach a precise physical meaning
to combinations like $L_1-\frac{1}{2}L_2$, which we pretend to be able
to compute in the fundamental theory.

This brings us to the crux of the matter. When we compute in chiral
perturbation theory and obtain a quantity that is ultraviolet finite, and,
in principle observable, is this quantity regulator independent or not?
If we were here dealing with a renormalizable theory,
the answer should be positively yes, but, of course, we are dealing here
with non-renormalizable quantum field theories, with a very pathological
ultraviolet behaviour as we have already mentioned.
Another way of phrasing the question is whether
renormalization-group invariants in chiral perturbation theory
may depend on the way we regulate the theory. We will see in the next
sections
that chiral invariance plays a very crucial and subtle role in settling this
issue in a very satisfatory manner. In
principle the result, even for finite quantities, can be completely
arbitrary.
We will illustrate this in the particular example of $\pi\pi$ scattering,
which we will discuss in detail in section 3. We will work in a rather
general class of regulators
in position space. In section 4 we will see how the potential
ambiguities
are resolved by demanding chiral invariance and that we can attach a
definite physical meaning to a combination such as (\ref{combi}). We
will propose
a general rule that regulators preserving chiral invariance must fulfil. In
section 5 we will extend the analysis to an arbitrary amplitude at one loop
and we will see that the analysis of section 3 carries over to any one-loop
process. Section 2 is devoted to general considerations about chiral
perturbation theory.
We will present our conclusions in section 6. In the Appendix
we analyze, rather exhaustively, several regulators that may be
of use in chiral perturbation theory.

Of particular interest is to implement the above ideas in the
symmetry
breaking sector of the Standard Model. Compiting alternatives for the scalar
sector of the Standard Model should lead to different values for the
appropriate coupling constants $L_i$\cite{EsHe}. We will not discuss this
point in the present work, although we believe that our results will be of
interest there. We note that a good theoretical control of the coefficients
$L_i$ is of particular importance.
The reason is twofold. First, the Electroweak Theory
contains some interactions that explicitly break the global
$SU(2)\times SU(2)$ symmetry, the number of possible counterterms is
greatly augmented (up to eight new operators can be written
up to ${\cal O}(p^4)$, and this  after
restricting ourselves to the $CP$-even sector\cite{Longhi,Appel1}).
Secondly,
in contrast to pion physics, we have precious little experimental information
concerning the scattering of longitudinal vector bosons, and
we will, most likely, remain in the same position for some time.

Finally, an additional motivation to undertake this work has been the
following. Technically, in chiral perturbation theory it is difficult
to go beyond one loop. Yet, many interesting properties require such
an analysis. Can we find regulators in which perturbation theory
is more manageable? Obviously, for that we need to know, first of all,
which is the most general regulator compatible with chiral symmetry.
An interesting computational scheme we have analyzed is the so-called
differential renormalization\cite{Lati}. We will discuss in one of the
appendices its advantages and shortcomings.

To summarize, we have tried to give a comprehensive and complete
account of the potential regularization ambiguities in non-linear sigma
models at the one loop level. We will start with a brief summary of the
basic
building blocks of chiral perturbation theory. Since there are excellent
references on this subject we will collect only the specific
points that we will need in later developments.

\section{Measure, Tadpoles and Counterterms}

The non-linear sigma model can be described in euclidean space-time by the
following partition function
\be
\label{pf}
Z=\int d\mu (U)  e^{- S(U)}
\ee
Minkowskian amplitudes will be obtained, as usual, by performing a Wick
rotation. As a rule, we will perform our calculations in euclidean space,
but present our results for the amplitudes in Minkowski space-time. To make
sense of the above expression we must, first of all,
determine which is the correct measure to use. Then we must introduce
a regulator in order to cut-off the high energy modes in the path-integral
and define (\ref{pf}) properly. In fact the two points are not
unrelated.

The obvious requirement that the measure of the path integral must fulfil
is, of course, chiral invariance. Both the action and the measure must
be invariant under $U(x)\to L U(x) R^\dagger$, with $L,R$ constant matrices
of $SU(N)$. $d\mu (U)$ must
therefore be an invariant group measure. The most convenient parametrization
of $U$,
the field belonging
to the group $SU(N)_L\times SU(N)_R/SU(N)_V$  describing the Goldstone
boson excitations,  is given by $U=\exp{ 2 i
{\pi^{a} T^{a}}/{f_{\pi}}} $ with $\pi^{a}$
being the pion fields and $T^{a}$ a properly normalized set of hermitian
generators of $SU(N)$.
The Goldstone boson fields ---pions, for short--- act as coordinates
in the group. In term of these coordinates we can construct the
group metric\cite{Zinn}
\be
g_{ab}=\delta_{ab}+{1\over {3 f_\pi^2}} \pi^c\pi^d(\delta_{ad}\delta_{cb}
-\delta_{ab}\delta_{cd})+{2\over {45 f_\pi^4}} \pi^c\pi^d\pi^e\pi^f(
\delta_{cd}\delta_{ab}\delta_{ef}-\delta_{cd}\delta_{ae}\delta_{bf})+...
\ee
The proper measure is then\cite{Zinn}
\be
d\mu (U)= \sqrt{\det g}\, \prod d\pi
\ee
The prefactor $\sqrt{\det g}$ can be exponentiated as
\be
\sqrt{\det g} = \exp{{1\over 2}\delta^{(4)}(0) Tr \ln{g}}
\ee
Expanding the trace in the previous expression gives rise to a series
of terms that, unlike $S(U)$, do not contain any derivatives. By
themselves
they are not chiral invariant, but they are required to compensate the
lack of chiral invariance of the ``flat" measure $\prod d\pi$. Of
course $ \delta^{(4)}(0)$ does not make much sense; it must be regulated
in the same way as the rest of the ultraviolet divergences that appear
in the perturbative expansion of $S(U)$. Even before doing that we
can already see that the pieces proportional to $\delta^{(4)}(0)$
originating from the measure cancel part of the tadpoles that
are generated in chiral perturbation theory. For instance, the
simplest tadpole, the one with two external legs, gives (we particularize
to $N=2$)
\be
\label{delta1}
\delta^{ab}{1\over f_\pi^2}({2\over 3}\delta^{(4)}(0) +{2\over 3}p^2
\Delta(0)) \ee
Where we have used that $\Box \Delta(x)=-\delta^{(4)}(x)$. $\Delta(x)$
is the pion propagator carrying a momentum $p^\mu$ and $a,b,...$
are $SU(2)$ indices. The contribution to this Green function originating
 from the measure is, at this order,
\be
\delta^{ab}{1\over f_\pi^2}(-{2\over 3})\delta^{(4)}(0)
\ee
thus cancelling the piece proportional to $\delta^{(4)}(0)$
in (\ref{delta1}). The next term,
of ${\cal O}(1/f_\pi^4)$, in the expansion of the measure is
\be
(\delta^{ab}\delta^{cd}+\delta^{ac}\delta^{bd}+\delta^{ad}\delta^{cb})
{1\over f_\pi^4}(-{4\over 45})\delta^{(4)}(0)
\ee
and adds to the  tadpole with four external legs, which on shell is equal to
\bea
\label{delta2}
&&(\delta^{ab}\delta^{cd}+\delta^{ac}\delta^{bd}+\delta^{ad}\delta^{cb})
{1\over f_\pi^4}(-{16\over 45})\delta^{(4)}(0)+\nn
\\
&&(\delta^{ab}\delta^{cd} s+\delta^{ac}\delta^{bd}
t+\delta^{ad}\delta^{cb} u) {1\over f_\pi^4}({10\over 9})\Delta(0)
\eea
We have introduced the usual kinematical invariants
\be
 s=(p_{1}+p_{2})^{2} \qquad  t=(p_{1}+p_{3})^{2} \qquad u=(p_{1}+
p_{4})^{2}
\ee
$a,b,c$ and $d$ are the $SU(2)$ indices carried by the pions with
momentum $p_1,p_2,p_3$ and $p_4$, respectively.
Both in (\ref{delta1}) and (\ref{delta2}) the $\delta^{(4)}(0)$'s
originate from the laplacian acting on the propagator, $\Box
\Delta(0)=-\delta^{(4)}(0)$. It is natural to identify the
$\delta^{(4)}(0)$ from the measure with $-\Box\Delta(0)$.
 When we regulate the short
distance singularities on the theory with the help of some
dimensional cut-off
$\eps$, $\Box\Delta(0)$ will be finite and proportional to $1/\eps ^4$
and so will be $\delta^{(4)}(0)$.

Apart from measure-induced terms and tadpoles
there are many more contributions to the four point amplitude. In general,
some additional pieces proportional
to $1/\eps ^4$ will be generated.(They will be discussed in detail in
the next section after the evaluation of the diagrams in fig. 1.)
Therefore we have contributions
proportional to $1/\eps ^4$ from very different sources: from the measure,
 from tadpoles and from ``normal" Feynman diagrams. Chiral invariance,
however dictates  that the net result has to be zero, because a piece of
the form $1/\eps^4$ cannot be absorbed by any chirally invariant
counterterm. In dimensional regularization, or in any other similar method
which automatically sets to zero all non-logarithmic divergences, the
above requirement is fulfilled by construction.
If a given regulator does lead to $1/\eps^4$
divergences it must be supplemented with appropriate counterterms to
bring the results into agreement with chiral invariance. This will
be our general philosophy. We shall write different conditions that
a regulator must comply with. If a given prescription does not lead
to results in agreement with the Ward identities of the theory, we will
supplement it with suitable counterterms. We will see in detail how
this procedure works for $\pi\pi$ scattering and then we will extend it
to an arbitrary one-loop process.

The next type of divergences we have to worry about are the quadratic
ones. Some of them are innocuous from the point of view of
preserving the Ward identities---they can be absorbed by a redefinition
of $f_{\pi}$. This is, for instance, the case of the two-legged tadpole
(eq. (\ref{delta1})). Redefining
\be
\label{redefi}
{1 \over f_\pi^2} \to {1\over f_\pi^2}-{1\over f_\pi^4}({2\over 3})
\Delta(0)
\ee
eliminates this term. Once regulated $\Delta(0)$ will be proportional
to $1/\eps ^2$. Chiral invariance implies that this same redefinition must
eliminate all the pieces proportional to $1/\eps ^2$ in all amplitudes.
For instance an amplitude with four external legs will contain terms
in $1/\eps ^2$ from tadpoles (eq. (\ref{delta2})) and from ``normal"
Feynman diagrams (fig. 1, see next section). Chiral
invariance requires those to add-up in such a way that they can be absorbed
by the redefinition (\ref{redefi}). Quadratic divergences that
cannot be eliminated
via a redefinition of $f_\pi$ must be absent. As in the case of cuartic
divergences, we can take the pragmatic attitude of using an arbitrary
regulator supplemented with suitable counterterms.

It turns out that from a computational point of view it is highly
advantageous to enforce the condition $\Delta(0)=0$ from the start.
While our analysis will go through without this restriction,
demanding
$\Delta(0)=0$ will simplify the calculations to a large extent.
Therefore we shall assume from now on that our regulator is such that
(perhaps
with the help of a counterterm) $\Delta(0)=0$. We will see in the appendix
how relaxing this condition does not change the results for the physically
relevant part of the amplitude.

\section{Non-linear $\sigma$ model to ${\cal O } (p^{4})$:  $\pi\pi$
scattering}

We shall use the $\pi\pi$ scattering amplitude as a battleground to
analyze in detail the regularization and renormalization ambiguities.
We will consider the $SU(2)$ case for simplicity and use the notation
and language of pion physics, although the conclusions are, obviously,
more general.
We will use the following notation to characterize $\pi\pi$
scattering
\be
\label{proc}
\pi^{a}(p_{1})+\pi^{b}(p_{2})\quad \longrightarrow
\quad\pi^{c}(p_{3})+ \pi^{d}(p_{4})
\ee
where $p^{i}$, $ i=1,..,4$ are the four momenta of the pions,
and $a,b,c=1,2,3 $ are isospin indices. The three Goldstone bosons form a
$I=1$ representation of the isospin group $SU(2)$. We will ignore
throughout the additional contributions that one should
include if we would take $SU(3)$ as the chiral symmetry group (as it is the
case in the real world). We will also set all the masses of the Goldstone
bosons to zero. Including the contributions from the kaons and mass
corrections changes nothing at the conceptual level, but makes all
calculations a lot more involved. There is an additional reason to stick
to the massless $SU(2)$ case. As
discussed in the introduction, the pions are here generic representatives of
the Goldstone bosons generated in the spontaneous breaking of the
$SU(2)_L\times SU(2)_R$ to its diagonal subgroup $SU(2)_V$. By just
replacing $f_\pi\simeq 93$ MeV by $v\simeq 250$ GeV one is describing the
scalar sector of the Standard Model of electroweak
interactions\cite{EqTh}.
The symmetry group there is indeed $SU(2)$ and, in addition, mass terms for
the Goldstone bosons are explicitly ruled out by gauge invariance.

The amplitude for this process can be written in a form that shows
explicitly the isospin structure\cite{man}. In terms of the Mandelstam
variables \be
\label{amp}
{\cal A}=F(s,t,u)\delta^{ab}\delta^{cd}+F(u,t,s)\delta^{ac}\delta^{bd}+
F(t,s,u) \delta^{ad}\delta^{bc}
\ee

In the effective chiral lagrangian, perturbation theory is organized
as an expansion in inverse powers of $f_\pi$, the pion decay constant.
Each pion field is accompannied by one of such powers. One loop
corrections to $\pi\pi$ scattering are thus down by a factor $1/f_\pi^2$
with respect
to tree level results. Since, on dimensional grounds, all operators
contributing to the effective lagrangian must be of dimension four, it is
clear that operators of ${\cal O}(p^4)$ like (\ref{L4}) have
coefficients in front of them which are down by a
factor $1/f_\pi^2$ with respect to the only operator of ${\cal O}(p^2)$,
(\ref{L2}). Thus to a given process, in particular $\pi\pi$
scattering, and up to
one-loop order in chiral perturbation theory we must therefore consider
several type of contributions.
On the one hand, there are contributions from the
${\cal L}^{(2)}$ lagrangian, both
at tree and one-loop level. In addition, the ${\cal
L}^{(4)}$
lagrangian contributes only at tree level. Altogether, the Feynman
diagrams that
contribute to the process (\ref{proc}) up to one-loop order are
depicted in fig. 1.

In the
absence of any external fields and mass terms, ${\cal L}^{(2)}$
and ${\cal L}^{(4)}$ for $SU(2)$ reduce to the following operators
\bea
\label{lagr}
 & &{\cal L}^{(2)}=\frac{f^2_{\pi}}{4} Tr(\partial_{\mu}U^{\dag}
\partial^{\mu}U) \nn \\
 & &{\cal L}^{(4)}= L_{1} Tr(\partial_{\mu}U^{\dag}
\partial^{\mu}U)^2+
 L_{2} Tr(\partial_{\mu}U^{\dag}
\partial_{\nu}U) Tr(\partial^{\mu}U^{\dag} \partial^{\nu}U)
\eea
If we now expand $U$ in powers of $1/f_\pi$ in (\ref{lagr}) one ends up
with
\bea
\label{eff}
 {\cal L}^{(2)}=&+&\frac{1}{2} \partial _{\mu} \pi^{i} \partial
^{\mu} \pi^{i}+
 \frac{1}{6 f^{2}_{\pi}} \partial _{\mu} \pi^{i} \partial ^{\mu}
\pi^{k}
 \pi^{j} \pi^{l} ( \delta^{il} \delta^{jk}-\delta^{ik} \delta^{jl})+
\nn \\
&+&\frac{1}{45 f^4_{\pi}} \pi^{m} \pi^{n} \pi^{j} \pi^{k}
\partial_{\mu} \pi^{r} \partial^{\mu} \pi^{s} (\delta^{mn}\delta^{rs}
\delta ^{jk}-\delta^{mn}\delta^{jr}\delta^{ks})+ ...
 \nn\\
 {\cal L}^{(4)}=&+& \frac{4 L_{1}}{f^4_{\pi}} \partial _{\mu} \pi^{i}
\partial^{\mu} \pi^{i} \partial _{\nu} \pi^{j} \partial ^{\nu}
\pi^{j}+
 \frac{4L_{2}}{f^{4}_{\pi}} \partial _{\mu} \pi^{i} \partial _{\nu}
\pi^{i} \partial ^{\mu} \pi^{j} \partial ^{\nu} \pi^{j}+ ...
\eea
where it is apparent that ${\cal L}^{(2)}$ contains, in addition to a free
kinetic piece,
an interaction term. Upon iteration we
will generate terms of ${\cal O}(1/f_\pi^4)$ and then we need to
include ${\cal L}^{(4)}$ too, if only for that reason. Of course we need
to include higher dimensional operators anyway because we are dealing
with a non-linear theory, plagued with ultraviolet divergences.
These divergences  cannot be reabsorbed by a
renormalization of the only coupling constant
we have at our disposal in  ${\cal L}^{(2)}$, namely $f_\pi$. We need
additional counterterms. How will they look? Well, if
we use a regulator that preserves the symmetry of the problem ---which we
certainly want to keep--- counterterms at the one loop level, on
locality, dimensionality and symmetry grounds will have to be necessarily of
the same form as the operators contained in ${\cal L}^{(4)}$.
Therefore
constants like $L_1$ and $L_2$ will, in general, require an infinite
logarithmic renormalization to make observables finite and to make
contact with experiment.

All this is, of course, well known and discussed in
textbooks on the subject\cite{Georgi}. However, the discussion is usually
left at this point, while our objective here is, for the reasons
mentioned
in the introduction, to answer the following two questions. Which is the
most general
regulator that preserves chiral invariance? And then, what is universal
in a general one-loop amplitude and what is regulator dependent?

We will thus be interested in calculating $F(s,t,u)$ in the most general
way. The contributions from the effective lagrangian (\ref{eff}) to
$F$ can be grouped in three parts
\be F(s,t,u)=F^{tree}(s,t,u)+F^{A}(s,t,u)+F^{B}(s,t,u)
\ee
$F^{tree}$ contains the tree level contribution to the
amplitude, both from  the ${\cal L}^{(2)}$ and  ${\cal L}^{(4)}$
pieces of the effective lagrangians. Explicitly,
\be
\label{tree}
F^{tree}(s,t,u)=\frac{s}{f_{\pi}^2}+\frac{8}{f_\pi^{4}}( s^{2}
L_{1}+(t^{2}+u^{2}) {L_{2}\over 2})
\ee
The one-loop contribution from ${\cal L}^{(2)}$ to the $\pi \pi$ scattering
amplitude has been separated into two pieces, $F^A$ and $F^B$, for reasons
that will
be explained below. $F^{A}_{2}(s,t,u)$ is given by
\be
\label{fa}
\frac{1}{2 f^{4}_\pi}\int d^{4}z\{s^{2}
\tg^{2}
e^{iz(p_{3}+p_{4})} + 4(p^{\nu}_{1} p^{\mu}_{2}+p^{\nu}_{4} p^{\mu}_{3})
\tg \partial_{\mu\nu}\tg(e^{iz(p_{2}+p_{4})}+e^{iz(p_{2}+p_{3})})\}
\ee
It turns out that all logarithmic divergences are in
$F^A$ and none is in
$F^{B}(s,t,u)$, which includes the rest of the one-loop amplitude.
Explicitly,
\bea
\label{caixes}
& &F^{B}(s,t,u)=\frac{1}{2 f^{4}_\pi}\{\int d^{4}z
(\frac{2}{9}(\Box \tg \Box \tg)+\frac{2}{9} \tg \Box^{2}
\tg+\frac{4}{3} s \tg \Box \tg) e^{iz(p_{3}+p_{4})}
\nn \\
& &\hspace{32mm}+(\frac{10}{9}(\Box \tg \Box \tg)-\frac{8}{9} \tg
\Box^{2} \tg+\frac{2}{3} t \tg \Box \tg) e^{iz(p_{2}+p_{4})}
\nn \\
&
&\hspace{32mm}+(\frac{10}{9}(\Box\tg\Box\tg)-
\frac{8}{9}\tg\Box^{2}\tg+\frac{2}{3}
u \tg\Box\tg) e^{iz(p_{2}+p_{3})} \}
\eea
Notice that $F^B$ contains {\it only} propagators and laplacian
operators acting on propagators, any of the contributions containing
at least a laplacian.
We have arrived at this unique decomposition
separating the integrals which can produce a laplacian acting on
a propagator inside the integral from the ones which cannot. In order to
do that one has to use identities like

$$\tg\Box\tg=\frac{1}{2}\Box\tg^2-\partial_{\mu}\tg\partial_{\mu}\tg$$

$$\tg\Box^2\tg=\frac{1}{2}\Box^2\tg^2-\Box\tg\Box\tg-4\partial_{\mu}\tg
\partial_{\mu}\Box\tg-2\partial_{\mu\nu}\tg\partial_{\mu\nu}\tg$$
\be
\tg\partial_{\mu\nu}\tg=\frac{1}{2}\partial_{\mu\nu}\tg^2-(\partial_{\mu
}\tg)(\partial_{\nu}\tg)
\ee

In all these manipulations it is assumed that the propagator $\tg(z)$
is regulated in some unspecified way to make it sufficiently regular at
short distances so that boundary terms can be safely neglected. Being
more precise,
let us now calculate $F^{A}(s,t,u)$ using a regulated
propagator defined in the following way
\be
\label{general}
\tg(z)=\frac{1}{4\pi^2}\int_{0}^{\infty} e^{-t z^{2}}
f^{\epsilon}(t) dt.
\ee
There are some obvious requirements that
$f^{\epsilon}(t)$ must fulfil. One has just been mentioned: the propagator
must be regular (for a finite cut-off) at short distances, giving rise
to well defined integrals (albeit dependent on
the cut-off and on the form of the regulating function $f^\eps$).
Furthermore,
by removing the cut off one must recover the usual free propagator
for a scalar particle
\be \label{prop}
\lim_{\epsilon \rightarrow 0} f^{\eps}(t)=1 \quad \Rightarrow
\quad \tg(z) \stackrel{\eps \rightarrow 0}{\rightarrow}
\frac{1}{4\pi^{2} z^{2}}
\ee
On dimensional grounds, $f^\eps (t)= f(\eps^2 t)$.
If we now introduce the expression for the propagator (\ref{general})
into
$F^{A}(s,t,u)$ and $F^B(s,t,u)$ everything can be written, obviously,
in terms of integrals of the regulating function $f^\eps$. For instance,
\bea \label{expresf}
& F^{A}(s,t,u)=& \frac{1}{32\pi^2 f^4}\{s^2 I_{1}^{\eps}(s)+t^2
I_{1}^{\eps}(t)+u^2 I_{1}^{\eps}(u)+ \\
& &(8 t^2 \deriv{}{t}-8 s)(\frac{1}{2}I_{2}^{\eps}(t)-I_{3}^{\eps}(t))
+(8 u^2 \deriv{}{u}-8 s)(\frac{1}{2}I_{2}^{\eps}(u)-I_{3}^{\eps}(u))\}
\nn \eea
where
\bea
\label{integ1}
& &I_{1}^{\eps}(s)=\int_{0}^{\infty}\int_{0}^{\infty} dt
dt'f^{\eps}(t) f^{\eps}(t') \frac{1}{(t+t')^{2}} e^{-s/(4 (t+t'))}
\nn \\
& &I_{2}^{\eps}(s)=\int_{0}^{\infty}\int_{0}^{\infty} dt dt'f^{\eps}(t)
f^{\eps}(t') \frac{1}{(t+t')} e^{-s/(4 (t+t'))}
\nn \\
& &I_{3}^{\eps}(s)=\int_{0}^{\infty}\int_{0}^{\infty} dt
dt'f^{\eps}(t) f^{\eps}(t') \frac{t^{2}}{(t+t')^{3}} e^{-s/(4 (t+t'))}
\eea
These integrals are all divergent when the cut-off is removed.
$I_1^{\eps}$ is logarithmically divergent and
$I_2^{\eps}$ and $I_3^{\eps}$ diverge quadratically as well as
logarithmically.
All three are dominated by an end-point singularity, which fixes the
coefficient of the log in an unique manner independently of the
detailed shape of the function $f^{\eps}(t)$ (see Appendix A).
Here we, of course, recover a familiar result in field theory, namely
the coefficient of the logarithmic singularity is universal, independent
of the regulating function $f^\eps$. This is a well known result in
renormalizable theories, like QED, where the logarithmic singularity is
the dominant one. It is much less clear, but it is nevertheless true,
in a theory like this one where the leading divergences are quadratic
(there may even exist quartic divergences, depending on the type of
regulator, as we will see in a moment).

We can write
\bea
\label{integ2}
& &I_{1}^{\eps}(s)=-\log{s \eps^{2}} + f + {\cal O}(\eps^2)
\nn \\
& &I_{2}^{\eps}(s)=\frac{s}{4} \log{s \eps^{2}} +
\frac{k_{1}}{\eps^{2}} + c s +{\cal O}(\eps^2)
\nn \\
& &I_{3}^{\eps}(s)=\frac{s}{12} \log{s \eps^{2}} +
\frac{k_{2}}{\eps^{2}} + d s + {\cal O}(\eps^2)
\eea
Except for the logarithmic coefficients all the other quantities
$(f,k_{1},k_{2},c,d)$
depend, of course, on the explicit form of $f^{\eps}(t)$.
In fact not all these coefficients are independent. There exist a
relation between $f$ and $c$  that can be found easily from the fact that
$ \deriv{I_{2}}{s} =-\frac{1}{4}I_{1}$,
\be
f=-4 c -1
\ee

By inserting  $I_1$, $I_2$ and $I_3$ into $F^{A}$ one gets (already
in Minkowski space)
\bea
\label{pbet1}
& F^{A}(s,t,u)=&-\frac{1}{96 \pi^2 f^4_{\pi}} \{ 3 s^{2} (\log{-s
\eps^{2}}-\beta_{1})
\nn \\
& &+t(t-u)(\log{-t \eps^2}-\beta_{2})+u(u-t)(\log{-u \eps^2}-\beta_{2})
\nn \\
& &+24 \frac{s}{\eps^2} (2 k_{2}-k_{1})\}
\eea
where
\be
\label{fet2}
\beta_{1}=-\frac{4}{3} -12 d  \quad\quad\quad
\beta_{2}=-1 -12 d
\ee
The last quadratic divergent term in (\ref{pbet1}) can be combined
with the four-legged tadpole and absorbed in a redefinition of $f_\pi$,
as discussed in section 2. This would uniquely fix the combination
$2k_2-k_1$ in a given scheme. (The general form of $F^{A}$ was
originally given by Lehmann\cite{Leh}.)

In order to complete the expression for the amplitude we still need to
compute
$F^{B}(s,t,u)$. First of all, it is easy to prove that $F^{B}$ does
not contain any log divergence. From dimensional
arguments we see that $F^B$ contains, potentially at least, quartic
divergences which are totally forbidden by chiral symmetry since it is
impossible to write a chirally invariant counterterm without derivatives
(as it would be required to absorb a $1/\eps^4$ divergence, on dimensional
grounds). So chiral symmetry has something to say on $F^B$. In fact it is
not difficult to see that in dimensional regularization, which complies with
the chiral Ward identities,
$F^B \equiv 0$. Indeed, in dimensional regularization
$ \Box \tg (x) = -\delta^{(n)}(x) $ and substituting this result into
(\ref{caixes}) we get
\be
F^B(s,t,u)= {2\over {9 f_\pi^4}}( 2 \delta^{(n)}(0) +s \Delta (0))
\ee
which is zero since in dimensional regularization
\be
\tg (0) =\int {d^nk\over{(2\pi)^n}} {1 \over  k^2}=0\qquad
\delta^{(n)}(0)= \int {d^nk\over{(2\pi)^n}} 1 = 0
\ee
Note that the coefficient of $\delta(0)$ is such that, combined
with the tadpole and the contribution from the measure, cancels out. This
is as it should be. The fact that $F^B=0$ in dimensional regularization
is not an accident, rather we
will see in the next section that the requirement that our regulator
satisfies the Ward identities is enough to guarantee that the
contribution from $F^B$ to the scattering amplitude vanishes when the
cut-off is removed. Let us then set $F^B=0$. The
complete amplitude reduces then to $F^{tree}+F^{A}$.

One can decompose the
amplitude (\ref{amp}) into three isospin channels $I=0$, 1 and 2. These
amplitudes with well defined isospin are given by the following
combinations
\be
\begin{array}{l}
T(0)=3 F(s,t,u)+F(u,t,s)+F(t,s,u)
\\
T(1)=F(t,s,u)-F(u,t,s)
\\
T(2)=F(t,s,u)+F(u,t,s)
\end{array}
\ee
Let's see what is the contribution from $F^{tree}$ (including the
${\cal O}(p^4)$ contribution) and $F^A$ to $T(0)$ and $T(2)$
\bea
\label{kbet1}
&T(0)=&\frac{2s}{f_{\pi}^2}+\frac{1}{96 \pi^2 f^4_\pi} \{
48 \frac{s}{\eps^2} (k_{1}-2k_{2}) -12 s^{2}
\log{-s \eps^{2}}+(2tu-8t^2)\log{-t\eps^2}\nn \\
& &+(2tu-8u^2)\log{-u\eps^2}
-(t^2+u^2)(24+240d)-tu(26+240d)
\}
\nn \\
&&+\frac{16}{f_{\pi}^4}\{L_{1}(2(t^2+u^2)+3tu)+\frac{L_{2}}{2}(3(t^2
+u^2)+2tu)\} \eea
\bea
\label{bet2}
&T(2)=&-\frac{s}{f_{\pi}^2}+\frac{1}{96 \pi^2 f^4_\pi} \{
-24 \frac{s}{\eps^2} (k_{1}-2k_{2}) -3 s^{2}
\log{-s \eps^{2}}+(ts-4t^2)\log{-t\eps^2} \nn \\
& &+(us-4u^2)\log{-u\eps^2}
-(t^2+u^2)(9+96d)-tu(8+96d)
\}\nn\\
&&+\frac{8}{f_{\pi}^4}\{L_{1}(t^2+u^2)+\frac{L_{2}}{2}(3(t^2+u^2)+4tu)\}
\eea
It is well known (and it can easily be
checked from the above expression)
that the contribution from $F^A$ to both $T(0)$ and $T(2)$ is ultraviolet
divergent, while the contribution from the one-loop chiral diagrams
to $T(1)$ is, in fact, finite. The divergences that appear in $T(0)$
and $T(2)$ can be removed by the following subtractions
\be
\label{Li}
L_1\to L_1=L_1(\mu) + \frac{1}{32\pi^2} (\frac{1}{12}) \log{\eps^2\mu^2}
\quad L_2\to L_2=L_2(\mu) + \frac{1}{32\pi^2} (\frac{1}{6})
\log{\eps^2\mu^2} \ee
With this prescription we can then find the value of $L_1(\mu)$ and
$L_2(\mu)$ at the fixed
(but arbitrary) scale $\mu$ from the experimental value of, for instance, the
phase shifts in the $I=0$ and $I=2$ channels. Of course the values of
$L_1(\mu)$ and $L_2(\mu)$ will be totally dependent on the
regulator one has used by an arbitrary constant because
the loop parts of divergent
amplitudes do depend on the way the cut-off is introduced. In addition
we may, in principle, add any constant we want to the subtractions implied
by (\ref{Li}). This would amount to a change in the renormalization
prescription, implying a modification in the value of the renormalized
coefficients $L_i(\mu)$. Let's
now see what
are the consequences of this renormalization procedure on the $I=1$ channel.

If we add all the contributions to the $I=1$ isospin one channel from
$F(s,t,u)$ (tree level and one-loop) that remain after imposing
chiral symmetry invariance (\ref{tree}) and (\ref{pbet1}) one
arrives to the following finite expression
\be
\label{t1}
T(1)=\frac{t-u}{f^2_\pi}-\frac{(t-u)}{96\pi^2 f^{4}_\pi}\{s \log{-s} + t
\log{-t} +u \log{-u} -3 s(\beta-256 \pi^2(L_{1}-\frac{1}{2}L_{2}))\}
\ee
$\beta$ is defined as $\beta=\beta_{2}-\beta_{1}$ and its value from
(\ref{fet2}) is
\be
\beta=\frac{1}{3}
\ee
{\it independently} of the regulator one has chosen. This remarkable
result shows
that $\beta$ is a universal quantity provided the regulator preserves
chiral invariance. On the other hand, both $T(0)$ and $T(2)$ depend on
various combinations of the two finite numbers $\beta_1$ and $\beta_2$,
which do
depend on the function $f^{\eps}(t)$ one has chosen to regularize
the amplitude. $\beta$, and consequently, the one-loop contribution to
$T(1)$ is not only cut-off independent, but also regulator
independent. In order to obtain this result it has been crucial to impose
chiral symmetry invariance.

The $I=1$ amplitude is, of course, observable.
$T(1)$ can be expanded into partial waves and by using the
effective range approximation one can find for the phase shift in the
$I=1$ channel and $l=1$ wave (this is the only one one has to include
due to Bose statistics)
\be
\label{cot}
\cot(\delta^{I=1}_{l=1})=\frac{96 \pi
f^2_\pi}{s}-\frac{3}{\pi}(\beta-256
\pi^2 (L_{1}-\frac{L_{2}}{2})-\frac{1}{9})
\ee
The
quantity on the r.h.s has to be a scheme independent quantity. The first
term is the tree contribution and is non ambiguous. The second term
$\beta-256 \pi^2(L_{1}-\frac{1}{2}L_{2})$ should be, as a consequence,
a scheme independent quantity but as we have just shown $\beta$
is regulator independent so  $L_{1}(\mu)-{1\over 2}{L_{2}}(\mu)$
is regulator independent too.

Eq. (\ref{cot}) has exactly the same form written either in terms of the
renormalized $L_i(\mu)$ or the unrenormalized $L_i$ coefficients if
we choose the subtraction prescription (\ref{Li}). If we take a
renormalization prescription different from the one implied by
(\ref{Li}),
i.e. if we modify the finite part of the subtractions, $\beta_1$ and
$\beta_2$ are modified in such a way that the amplitudes $T(0)$ and
$T(2)$ remain numerically the same. The value of
$\beta=\beta_{2}-\beta_{1}$ also gets modified accordingly
although, obviously, numerically
(\ref{cot}) remains
the same. This is a zero-sum game, we may well choose to transfer
finite parts back and forth between $\beta$ and the renormalized value
of $L_{1}(\mu)-\frac{1}{2}L_{2}(\mu)$ if we so wish, but the fact is
that in any chiral invariant regularization scheme $\beta=1/3$ is an
unambiguous prediction of chiral perturbation theory and a comparison
between the theoretical and experimental value for $L_1-\frac{1}{2}L_2$
is a priori meaningful.

\section{Ward identities}

Let's now see how the chiral Ward identities impose severe constraints
on our regulator.
The use of a regulating function $f^{\eps}(t)$ for the propagator
eq. (\ref{general}) implies a modification of the momentum space
inverse propagator \be
p^2\to p^2 +p^2 {\cal F}(\eps^2 p^2)
\ee
The free part of the effective lagrangian ${\cal L}^{(2)}$ then
changes in the following way
\be
{\cal L}^{(2)}=\frac{1}{2} \pi^{i}(\Box+{\cal F}(\eps^2\Box)\Box) \pi^{i} +
... \ee
The dots stand for the (unmodified) interaction terms. The point to
analyze now is whether under a chiral transformation, the
action is still chiral invariant or not. The answer is that, in general,
is not.
Under a left chiral transformation ($L\neq 1, R=1$) the pion field
transforms as \be
\label{chiral}
\delta\pi^{a}(x)=\frac{f_\pi
\omega^{a}}{2}-\frac{1}{2}\eps^{abc}\omega^{b}\pi^{c}+\frac{1}{6f_\pi}(
\delta^{ab}\delta^{cd}-\delta^{ad}\delta^{bc})\omega^{d}\pi^{b}\pi^{c}+
{\cal O}(1/f^2_\pi)
\ee
For a right chiral transformation ($L=1, R\neq 1$) there is a change in the
sign of the second term on the r.h.s.

For an arbitrary ${\cal F}$ the action is not chiral invariant under this
transformation, but rather $\delta S\neq 0$. If we now require
chiral symmetry to
hold, it would imply either to include counterterms to compensate the
noninvariance, i.e. the non-zero value of $\delta S$  or conditions on
${\cal F}$ (or, equivalently, on $f^\eps (t)$).
Let's demand invariance at the level of $S$-matrix elements
\be
{\langle\delta S\; a^{\alpha_{1}\dag}(p_{1})...a^{
\alpha_{n}\dag}(p_n)\rangle}= 0
\ee
Since we are interested in $\pi\pi$ scattering at the one loop level, a
simple counting of powers of $f_\pi$ shows that
the relevant diagrams to analize to the order we are calculating are
the ones shown in fig. 2. They lead to the following matrix
element \be
\label{longint}
\int d^4x \int d^4y \langle 0|\delta\pi^{a}(x){\cal F}(\eps^2\Box) \Box
\pi^{a}(x) L_{int}(y) a^{\alpha\dag}(p_{1}) a^{\beta\dag}(p_{2})
a^{\gamma\dag}(p_{3}) |0\rangle
\ee
Adding and subtracting 1 to ${\cal F}(\eps^2\Box)$ and using that
\be
\Box(1+{\cal F}(\epsilon^2 \Box))\tg (x) = -\delta^{(4)}(x)
\ee
where $\tg$ is the {\it regulated} propagator, and imposing $\tg(0)=0$
(see
the discussion in section 2), one immediately sees that all terms will
necessarily contain laplacians acting on the propagators. In fact
one arrives exactly to the same type of integrals that appear
in (\ref{caixes}). If one expands them in order to separate the quartic
and quadratic divergences and the finite parts one gets
\bea
\label{coe1}
& &T_{1}(s)=\int d^4 z \Box \Delta \Box \Delta e^{iPz}=
\frac{A}{\eps^4}+\frac{B s}{\eps^2}+G_{1} s^{2} + {\cal O}(\eps^2)
\nn\\
& &T_{2}(s)=\int d^4z \Delta \Box^2 \Delta e^{iPz}=
\frac{C}{\eps^4}+\frac{D s}{\eps^2}+G_{2} s^{2} + {\cal O}(\eps^2)
\nn\\
& &T_{3}(s)=\int d^4z  \Delta \Box \Delta e^{iPz}=
\frac{E}{\eps^2}+G_{3} s + {\cal O}(\eps^2)
\eea
The contributions proportional to $\delta^{(4)}(0)$ from
(\ref{longint}) cancel
exactly the terms from the measure, so we do not need to worry about
them.

Notice that the previous decomposition does not include any logarithmic
divergence. That this is so, can be proved easily by introducing the
general propagator (\ref{general}) into (\ref{coe1}),

$$\hspace{-15mm} T_{1}=\int_{0}^{\infty}\int_{0}^{\infty} dt
dt'f^{\eps}(t) f^{\eps}(t') \frac{ e^{-s/(4(t+t'))}}{(t+t')^{6}}(s
t^4t'+\frac{s^2}{16}t^2t'^2+ 12(t^4t'^2+t^3t'^3))\hfill $$
$$T_{2}=\int_{0}^{\infty}\int_{0}^{\infty} dt
dt'f^{\eps}(t) f^{\eps}(t')
\frac{e^{-s/(4(t+t'))}}{(t+t')^{6}}(\frac{3s}{2}
(t^4t'+t^3t'^2)+\frac{s^2}{16}t^4
+12(t^4t'^2+t^3t'^3))$$
\be
\label{dec}
T_{3}=\frac{1}{4}\int_{0}^{\infty}\int_{0}^{\infty} dt
dt'f^{\eps}(t) f^{\eps}(t')
\frac{e^{-s/(4(t+t'))}}{(t+t')^{4}}(-4t't^2-\frac{st^2}{4}) \hfill
\ee
and following the analisis of appendix A one can see that
the
logs cancel.

Not all the coefficients in (\ref{coe1}) are free. There are some
relations amongst them
\bea
\label{eq}
&A=C \quad
&2D=2B-E
\eea
By introducing (\ref{coe1}) and (\ref{eq}) in the matrix element one ends
up with
\bea
& &\hspace{-18mm} \langle\delta S\,  a^{\alpha \dag}(p_{1}) a^{\beta
\dag}(p_{2}) a^{\gamma \dag}(p_{3})\rangle ={\cal O}(\frac{1}{f_{\pi}})-
\frac{1}{f^3_{\pi}} \frac{\omega^{b}}{36}\delta(\sum
p)\delta^{b\alpha}\delta^{\beta\gamma} \times
\{
\frac{8A}{\eps^4}+\nn\\
&\qquad &\frac{1}{\eps^2}[2p_{1}^{2}(2D+7E)+(2D-4E)(p_{2}^2+p_{3}^2)]+
\nn\\ &\qquad&(G_{1}+G_{2}+3G_{3})(2p_{1}^4+p_{2}^4+p_{3}^4)+
(9G_{1}-9G_{2}-3G_{3})(p_{1}^2 p_{3}^2+p_{1}^2 p_{2}^2)+\nn\\
&\qquad&p_{2}^2 p_{3}^2(-18 G_{1}+18G_{2}-2G_{3})+R(p_{i},\eps)+{\cal O}
(\eps^2) \}+ \nn\\
&\qquad &(\alpha \leftrightarrow \beta\quad   p_{1}
\leftrightarrow p_{2})+(\alpha \leftrightarrow \gamma\quad   p_{1}
\leftrightarrow p_{3} )
+{\cal O}(1/f_{\pi}^5) = 0
\eea
where $R(p_{i},\eps)$ collects the contribution that comes from the
action of $F(\eps^2 \Box)$ over the external legs, which (like the
simplifying assumption $\Delta(0)=0$) do not modify the conclusion for
the pieces of ${\cal O}(p^4)$. If we set
to zero all the different tensorial structures, we obtain a set of equations
for the coefficients $A,B,...$. It turns out that the only solution of
this system of equations turns out to be
\bea
& &A=C=0
\nn\\
& &B=D=E=0
\nn\\
& &G_{1}=G_{2}=G_{3}=0,
\eea
implying that, up to terms that vanish when the cut-off is removed,
$T_1=T_2=T_3=0$ if our regulator is to comply with chiral
invariance.
But these integrals are exactly the same appearing in $F^B$, so
 from the requirement of chiral invariance of the regulator we conclude
that $F^{B}(s,t,u)=0$.

 From the preceding analysis we can extract a set of rules that are
bound to yield results respecting chiral invariance in chiral perturbation
theory.

$\bullet$ Take the amplitude in position space, either calculating the
matrix element directly in position space or Fourier transforming from
momentum space.

$\bullet$ Separate all the integrals that can lead to a
laplacian
acting on a propagator from the ones that cannot before regularizing.

$\bullet$ Use $\Box \Delta(x) = -\delta(x)$ inside the integrals

$\bullet$ Integrate the delta functions and end up with
integrals that do not contain any $\Box$ inside, and terms like
$\Delta(0)$ and $\Box \Delta(0)$.

$\bullet$ Choose any regulator good enough for the remaining integrals
and such that verifies $\Delta(0)=0$ and $\Box
\Delta(0)=0$. For instance,
\be
\Delta(x)={1\over x^2}{(1-({x\over {a \eps}})^4
K_{1}({x\over {a\eps}})^{4})} \ee
If the regulator does not automatically cancel all tadpoles, we can
always remove them by suitable counterterms. This has no effect on the
relevant finite parts.

\section{A General One-loop Process }

In this section we will show that the previous procedure
can be extended to an arbitrary one-loop process with $m$ external
legs. That is,
we will see that in chiral perturbation theory, provided one
is respectful with the chiral invariance of the theory,
the loop part of finite amplitudes is
automatically independent of the regulator.

The most general diagram with $n$ vertices and $m$ external propagators that
one can construct from the lagrangian
\be
\label{int}
 {\cal L}^{(2)}=\frac{f^2_{\pi}}{4} Tr(\partial_{\mu}U^{\dag}
\partial^{\mu}U) \ee
has $n$ internal propagators, $2n$ derivatives and
$p^{i}=q^{i}+r^{i}+s^{i}+...$, $i=1,...n$ external momenta. $q^i$, $r^i$,
etc. label the different external momenta flowing into vertex $i$.
The
conservation rule $\sum_{i=1}^{n} p^{i}=0$ holds. A generic diagram is
shown in fig. 3.

We start our calculation in position space and label by $z_1$, ... $z_n$
the vertex coordinates.
Taking into account the way the derivatives act on the propagators,
using $x_{i}=z_{i}-z_{i+1}$ $i=1...n-1$ and integrating by parts
according
to the set of rules of section 3 one arrives to the following general
decomposition for any diagram, at one loop,
\be
\delta^{a_{1} a_{2}}\delta^{a_{3} a_{4}}...
\delta^{a_{m-1}
a_{m}} F^{(nm)}(q^{j},r^{j},...)+ {\rm permutations}
\ee
where we have to include all possible permutations
of isospin indices and momenta. The $F^{(nm)}$ have the
form
\bea
\label{gen}
& F^{(nm)}=&\frac{1}{f^{m}_\pi}\{\sum_{i=0}^{n} g_{(ni)}
^{(nm) {\mu_{1}...\mu_{i}}}(q_{1},r_{1},...)
I^{(n)}_{\mu_{1}...\mu_{i}}(p_{1},p_{2},...p_{n-1})
\\
& &+\sum_{k=2}^{n-1}\sum_{i=0}^{k} g_{(ki)}
^{(nm) \mu_{1}...\mu_{i}}(q_{1},r_{1},...)
I^{(k)}_{\mu_{1}...\mu_{i}}(p_{1}-p_{k},p_{2}-p_{k},...p_{k-1}-p_{k})\}
\nn
\eea
where
$I^{(s)}_{\mu_{1}...\mu_{i}}(p_{1},...,p_{s-1})$ is the integral
\be
\int
d^4 x_{1}...d^4 x_{s-1} e^{i(\sum_{j=1}^{s-1} p_{j}x_{j})}
\Delta(x_{1})...\Delta(x_{s-1})\partial_{\mu_{1}...\mu_{i}}\Delta(x_{1}+
...+x_{s-1}) \ee
Let us see how this decomposition works in a simple example.
Let's suppose that  we want to describe a process that involves 6 external
particles. This process
will receive contribution from diagrams with 3 and 2 vertices.
Let us concentrate only in the first
diagram for which $n=3$ and $m=6$ depicted in Fig. 4. Using the
interaction
lagrangian (\ref{int}), integrating by parts and using that $\Box
\Delta(x)=-\delta^{(4)}(x)$ the contribution from the diagram is, apart
 from isospin indices,

\bea
\displaystyle
F^{(3,6)}&=&\frac{1}{f^6_{\pi}}\{ g_{(3,0)}^{(3,6)}\int d^4x_{1}
d^4x_{2} e^{i(p_{1}x_{1}+p_{2}x_{2})}
\Delta(x_{1})\Delta(x_{2})\Delta(x_{1}+x_{2})
\nonumber\\
&+&g_{(3,1)}^{(3,6){\mu_{1}}}\int d^4x_{1} d^4x_{2}
e^{i(p_{1}x_{1}+p_{2}x_{2})}
\Delta(x_{1})\Delta(x_{2})\der{}{x_{1}^{\mu_{1}}}\Delta(x_{1}+x_{2})
\nn\\
&+&g_{(3,2)}^{(3,6){\mu_{1}\mu_{2}}}\int d^4x_{1} d^4x_{2}
e^{i(p_{1}x_{1}+p_{2}x_{2})}
\Delta(x_{1})\Delta(x_{2})\der{}{x_{1}^{\mu_{1}}}\der{}{x_{1}^{
\mu_{2}}}\Delta(x_{1}+x_{2})
\nn\\
&+&g_{(3,3)}^{(3,6){\mu_{1}\mu_{2}\mu_{3}}}\int d^4x_{1} d^4x_{2}
e^{i(p_{1}x_{1}+p_{2}x_{2})}
\Delta(x_{1})\Delta(x_{2})\der{}{x_{1}^{\mu_{1}}}\der{}{x_{1}^{
\mu_{2}}}\der{}{x_{1}^{\mu_{3}}}\Delta(x_{1}+x_{2})
\nn\\
&+&g_{(2,0)}^{(3,6)}\int
d^4x_{1} e^{i(p_{1}-p_{2})x_{1}} \Delta(x_{1})\Delta(x_{1})
\nn\\
&+&g_{(2,1)}^{(3,6){\mu_{1}}}\int d^4x_{1}
e^{i(p_{1}-p_{2})x_{1}}
\Delta(x_{1})\der{}{x_{1}^{\mu_{1}}} \Delta(x_{1})
\nn\\
&+&g_{(2,2)}^{(3,6){\mu_{1} \mu_{2}}}\int d^4x_{1}
e^{i(p_{1}-p_{2})x_{1}}
\Delta(x_{1})\der{}{x_{1}^{\mu_{1}}}\der{}{x_{1}^{\mu_{2}}}
\Delta(x_{1})\}
\nn\\
\eea
In the last three terms the equation $\Box \Delta (x)=-\delta^{(4)}(x)$
has been used once. Of course it is also easy to see that, after
neglecting terms like $\Delta(0)$ or $\Box\Delta(0)$, the $\pi\pi
\to \pi\pi$ amplitude, which then reduces to the $F^A$ term (eq.
\ref{fa}), is also of the generic form (\ref{gen}).

The decomposition given in (\ref{gen}) shows what type of integrals
appear in the calculation of a given diagram after the
application of the set of rules given in section 3. Of all these
integrals we will be only interested in those that by power counting
are potentially divergent, since the integrals that are convergent
by naive power counting are certainly independent of the regulator.
It is clear
 from the non-linear nature of
(\ref{int}) that a physical process with $m$
external particles gets contributions, at the one loop level,
 from diagrams with
$n=m/2$ internal lines all the way down to diagrams with $n=2$
(we do not take into account tadpole diagrams).
A given physical process with
$m$ external particles
receives contributions
 from the same type of integrals that appear for a process with
$m-2$, $m-4$, etc. external particles (all the way down to
4, again ignoring tadpoles). The question is: are there new
divergent integrals allowed by power counting for increasing values
of $m$ or does the appeareance of new types of divergent integrals stops at
some point?
Fortunately, the number of divergent integrals that appear is rather limited.
For instance,  the process with $m=4$ has (excluding tadpoles)
only one class of divergent diagram that leads to two
independent integrals.
A process with $m=6$ gets contribution
 from diagrams with $n=3$ and $n=2$ with four type of
divergent integrals, the two that contributed for $m=4$ plus two new ones.
For a process with $m=8$ legs diagrams with
$n=4,3$ and $2$ vertices contribute. Only one new divergent integral
appears, making a total of 5 different divergent integrals. For $m=10$
and beyond no new divergent integrals appear.

We are thus confronted with only 5 possible divergent integrals
for {\it any} process in chiral perturbation theory at the one
loop level.
\be
I^{(2)}(p_{1})=\int d^4x e^{iP_{1}x} \Delta(x)^2
\ee
\be
I^{(2)}_{\mu_{1}\mu_{2}}(p_{1})=\int d^4x e^{iP_{1}x}
\Delta(x) \partial_{\mu_{1}\mu_{2}} \Delta(x)
\ee
\be
\label{eq1}
I^{(3)}_{\mu_{1}\mu_{2}}(p_{1},p_{2})=\int d^4x_{1}
d^4x_{2}
e^{i(P_{1}x_{1}+P_{2}x_{2})}\Delta(x_{1}) \Delta(x_{2}) \partial
_{\mu_{1}\mu_{2}} \Delta(x_{1}+x_{2})
\ee
\be
\label{eq2}
I^{(3)}_{\mu_{1}\mu_{2}\mu_{3}}(p_{1},p_{2})=\int
d^4x_{1} d^4x_{2}
e^{i(P_{1}x_{1}+P_{2}x_{2})}\Delta(x_{1}) \Delta(x_{2}) \partial
_{\mu_{1}\mu_{2}\mu_{3}} \Delta(x_{1}+x_{2})
\ee
\be
\label{eq3}
I^{(4)}_{\mu_{1}\mu_{2}\mu_{3}\mu_{4}}(p_{1},p_{2},p_{3})=
\ee
$$
\int d^4x_{1} d^4x_{2} d^4x_{3}
e^{i(P_{1}x_{1}+P_{2}x_{2}+P_{3}x_{3})}\Delta(x_{1}) \Delta(x_{2})
\Delta(x_{3}) \partial
_{\mu_{1}\mu_{2}\mu_{3}\mu_{4}}\Delta(x_{1}+x_{2}+x_{3})$$
All of them contain only
logarithmic divergences except for the second one
which has an additional quadratic divergence.
If one expands (\ref{eq1}) (\ref{eq2})
and (\ref{eq3}) in terms
of all the possible tensorial structures, one sees that
all integrals can be split into a finite (by power counting) part,
which is non-ambiguous,  plus some terms that contain the
divergent contributions. In fact, all the divergences are
concentrated in only two
pieces $I^{(2)}$ and $I_{\mu \nu}^{(2)}$, which turn out to be
proportional to the integrals that appear in $\pi\pi$ scattering, namely
$I^{(2)}$ and $I^{(2)}_{\mu\nu}$

$$I^{(3)}_{\mu_{1}\mu_{2}}(p_{1},p_{2})=\frac{1}{4} g_{\mu_{1} \mu_{2}}
I^{(2)}(p_{1}-p_{2})+{\rm finite}$$
$$I^{(3)}_{\mu_{1}\mu_{2}\mu_{3}}(p_{1},p_{2})=-\frac{i}{12}\{(p_{1}+
p_{2})^{\mu_{3}}
g_{\mu_{1}\mu_{2}}+(p_{1}+p_{2})^{\mu_{2}}g_{\mu_{1}\mu_{3}}+
$$
$$\hspace{42mm}(p_{1}+p_{2})^{\mu_{1}}g_{\mu_{2}\mu_{3}}\}
I^{(2)}(p_{1}-p_{2})+{\rm finite}$$
\be
\label{ieq}
I^{(4)}_{\mu_{1}\mu_{2}\mu_{3}\mu_{4}}(p_{1},p_{2},p_{3})=\frac{1}{24}(
g_{\mu_{1}\mu_{2}}
g_{\mu_{3}\mu_{4}}+g_{\mu_{1}\mu_{3}}g_{\mu_{2}\mu_{4}}+g_{\mu_{1}\mu_{4}
 }g_{\mu_{2}\mu_{3}}) I^{(2)}(p_{1}-p_{2})+{\rm finite}
\ee
This implies that having regularized the one-loop process with two
vertices (section 3), where the $I^{(2)}$ and
$I^{(2)}_{\mu_{1}\mu_{2}}$ integrals appear, one has by the same
token regularized all the one-loop processes.
In a way this should not be too surprising; recall that all
logarithmic divergences at the one loop level can be eliminated by a
redefinition
of just two coupling constants of the ${\cal L}^{(4)}$ lagrangian.

For any amplitude ${\cal A}$ with $m$ external particles one can
construct a number of linear combinations of the different
$F^{(nm)}(s_{1},s_{2},s_{3},...)$
($s_{i}$
are the invariant quantities that one can construct with $2m$ independent
momenta) such that the chiral logs cancel.
Because it is
a finite quantity, if we call this combination ${\cal B}$, it obviously
verifies
\be   \mu {\partial\over{\partial \mu}} {\cal B} =0
\ee
$\mu$ being the renormalization scale. For the $\pi\pi$ scattering
amplitude, that we have
analyzed in previous sections, we would have $m=4$, all the amplitudes
can be expressed in terms of
$F^{(2,4)}(s,t,u)$ plus permutations of $s$, $t$ and $u$. The
combination
that leds to a finite amplitude is, obviously, $T(1)$, the $I=1$
channel amplitude (\ref{t1})
\be
{\cal B}=F^{(2,4)}(s_{2},s_{1},s_{3})-F^{(2,4)}(s_{3},s_{2},s_{1})
\ee
The functions  $F^{(nm)}$ contain divergent (and, hence, regulator
dependent) integrals as well as finite, unambiguos terms. From the
discussion leading to eq. (\ref{ieq}) we learn that all divergent
integrals can be reduced unambiguously to two
integrals $I^{(2)}$ and $I^{(2)}_{\mu_{1}\mu_{2}}$.
We can still go further and isolate the logarithmic divergences of
these integrals into one structure. Using a tensorial decomposition of
$I^{(2)}$ and $I^{(2)}_{\mu_{1}\mu_{2}}$ we can write,
\be
I^{(2)}=\int d^4x \Delta(x)^2 e^{iPx}=\frac{1}{16 \pi^2} B
\nn
\ee
\be
I^{(2)}_{\mu\nu}=\int d^4x \Delta(x) \partial_{\mu\nu}
\Delta(x) e^{iPx}=\frac{1}{16 \pi^2} [A(g_{\mu\nu} p^2-4 p_{\nu}
p_{\mu})+D g_{\mu\nu} \frac{1}{\eps^2} + C p_{\nu} p_{\mu}]
\ee
The expression for $A,B,C$  and $D$ can be deduced by using the
propagator (\ref{general}) and the integrals (\ref{integ1}) and
their solution (\ref{integ2})
\bea
\label{ab}
&&A=-\frac{1}{12} \log{s \eps^2}+2d-c
\nn\\
&&B=-\log{s \eps^2} -4c -1
\nn\\
&&C=12d -4c +\frac{1}{3}
\nn\\
&&D=2k_{2}-k_{1}
\eea
All the
arbitrariness in choosing one regulating function $f^{\eps}(t)$ or
another is encoded, except for a redefinition of $f_\pi$,  in the
coeficients $c$ and $d$. The key point is
that all the logarithmic divergences are concentrated in only one structure
for each integral, $A$ for $I^{(2)}$ and $B$ for $I^{(2)}_{\mu_{1}\mu_{2}}$.

The finite combination of the $F^{(n,m)}$ that defines the
corresponding ${\cal B}$ can be splitted into two pieces.
One contains all the finite contributions that come from the manifestly
convergent
integrals. Of course, as these integrals are well defined {\it per se};
they don't need to be regularized, so their value is fixed and
is nonambiguos (scheme independent). The second piece would contain
the finite parts that accompany the divergences of $A$ and $B$.
There  will exist only one combination of $A$ and $B$ that is
finite; from (\ref{ab}) one can see that it is $A-\frac{1}{12}B$.
This unique combination fixes the way the scheme dependent quantities
$c$ and $d$ appear in ${\cal B}$. Symbolically
\be
{\cal B}=\sum_{\alpha} C_{\alpha} F^{\alpha} [I^{(2)},I^{(2)}_{
\mu\nu},\;{\rm finite}]=\sum_{\alpha} C_{\alpha}
F^{\alpha}[A,B,C,\;{\rm finite}] \ee
where the index $\alpha$ represents all diagrams, including permutations
of indices, that contribute to the finite quantity ${\cal B}$. From
the previous discussion,
\bea
\label{last}
{\cal B}&=&{\rm finite}+(A-\frac{1}{12} B)(\;{\rm finite}) \nn\\
&=&{\rm finite}+\frac{1}{6}(12d -4c+\frac{1}{2})(\;{\rm finite})
\eea
where ``finite" means some quantity that can be written in terms of
convergent integrals alone---hence, unambiguous. All the non-universal
dependence on the type of regulator is potentially encoded in the
combination $12d-4c$.
But this particular structure $12d-4c$ is exactly the same one finds
 from $I_{\mu\nu}$ by contracting with the metric $g_{\mu\nu}$
\be
\int d^4x {\Delta\Box\Delta}
e^{iPx}=\frac{1}{16 \pi^2}[\frac{4}{\eps^2}(2k_{2}-k_{1})+
 p^2 (12d-4c+\frac{1}{3})]
\ee
In section 4 we have shown that this
integral has to be zero, on chiral invariance grounds, so then
\be
12d-4c=-\frac{1}{3}
\ee
and putting this result into (\ref{last}) one finally ends up with an
expression for ${\cal B}$ that does not contain any remnants of the
arbitrariness of the regulating function $f^{\eps}(t)$ one has
chosen
\be
{\cal B}={\rm finite}+\frac{1}{36} (\;{\rm finite})
\ee
Renormalization group invariants are automatically independent of the
regulator
in chiral perturbation theory, at least at the one loop level, provided
the regulator respects the chiral symmetry.

\section{Conclusions}

Chiral lagrangians provide a consistent framework to  describe the
interactions amongst Goldstone bosons. These are non-linear,
non-renormalizable
theories in four dimensions. To make sense of these theories and
to compare calculations and experiment we must, first of all,
absorb the rather severe ultraviolet divergences into effective
couplings. In spite of the non-renormalizable character of the theory
this can be accomplished order by order in a momentum expansion and the
renormalized effective coupling obtained from comparison with the
experimental data.
This program has been successfully applied to very many different physical
applications with very satisfactory results.

If one wants to be more ambitious and compare the effective couplings that
can be deduced from experiment with theoretical predictions, or, even if
for more formal  reasons one wants to somehow attach
more  field theoretical respectability to chiral perturbation theory,
one needs to know to what extent the results obtained in such
non-renormalizable theories may depend on the way the cut-off is introduced.
We have presented an in-depth study of these issues here.
We have found that, generally speaking, observables which are
renormalization-group invariants are completely independent
of the regulator, provided the latter respects the Ward identities
of the theory. We have proven that only at the one loop level, but
we trust it must hold at higher orders. Likewise, we have not considered
the addition of gauge fields to the chiral lagrangian, but we believe, too,
that this should not change matters.

As a conclusion, combinations of the ${\cal O}(p^4)$ coefficients
that are renormalization group invariant, have unambiguous,
finite values that can be extracted from experiment and compared
with predictions from some more fundamental theory to which the
chiral lagrangian is an approximation at long distances. This is
of particular importance in applications of chiral lagrangian techniques
to the symmetry breaking sector of the Standard Model to discern, for
instance, between a strongly interacting Higgs or more exotic possibilities.

Our approach has been to use a general regulator in position space
and demand the fulfilment of the chiral Ward identities. To simplify
matters we have demanded an extra condition for this propagator,
namely $\Delta(0)=0$, but this is only a technical point. Then we
isolate from the different amplitudes a part $A$ that is non-zero and
contains the relevant information and a part $B$ which must vanish
upon restricting ourselves to chirally invariant regulators. It matters
little where the regulator complies directly with the chiral Ward
identities or these have to be enforced by counterterms. This is only
a semantic distinction. We have noted, that counterterms to restore
chiral invariance have to be
considered in any case since they are generated by the measure anyway.
When we consider the non-zero $A$ part, a universal
result is obtained, even if the regulator we use is non-chirally
invariant. We can work out this $A$ part in the regulator we please.

Finally, we have collected a number of regulators, including
the recently proposed differential renormalization, and several technical
details in the appendices in the hope that this material can be of use
to the reader.

We hope to have clarified some of the issues raised in the introduction.

\section*{Acknowledgements}

\hspace{\parindent}
Discussions with E. de Rafael, A. Dobado, M.J. Herrero, P. Haagensen,
 H. Leutwyler, S. Peris and F.J. Yndur\'ain
are gratefully acknowledged. This work has been partially
supported by
CICYT project no.\ AEN90-0033 and by  CEE Science Twinning Grant
SCI-0337-C(A).
J. M. acknowledges a fellowship from Ministerio de Educaci\'on y Ciencia.

\clearpage

\appendix

\section{End Point Singularities}

In this appendix we will comment on the evaluation of the
integrals in (\ref{integ1}). These integrals contain all the
information
needed in order to determine the coefficients of the chiral
logarithms and the scheme independent quantity $\beta$.

The key point is that all three integrals have a universal dependence
either on $\log s$ or $s\log s$ that is easy to determine. Let us recall
the minimum requirements on the regulating function $f^\eps(t)$.
First, the function $f^{\eps}(t)$ must define a well-behaved
propagator everywhere (this implies, in particular, that $f(t) \sim
1/t^{(2+\alpha)}$, $\alpha \ge 0$), and second, when removing the cut-off
one must recover the usual propagator, forcing
$f(0)=1$.
It is useful to perform a change of variables on these integrals.
\be
v=t+t' \quad\quad\quad u=t
\ee
The $I_{1}^{\eps}$ integral, in particular, having rescaled the cut off
$\eps$, reads
\be
I_{1}^{\eps}=\int_{0}^{\infty}dv e^{-\frac{s \eps^2}{4 v}}\frac{1}{v^2}
g(v) \ee
where $g(v)=\int_{0}^{v}du f(u) f(v-u)$.
 From the requirements on
$f^{ \eps}(t)$
\be
g(v)\relstack{\sim}{v \rightarrow \infty} 0 \quad\quad\quad\quad
g(v)\relstack{\sim}{v \rightarrow 0} v
\ee
If we set $\eps=0$  we find a logarithmically divergent
integral dominated by the singularity at $v=0$. We can
split the range of integration in two, from $0$ to a certain
value $c$ and from $c$ to $\infty$. The last one will be convergent,
while in the first one we can approximate $g(v)\sim v$ if we choose
$c$ to be small enough. Then
\be
I_{1}^{\eps}(s)=-\log{\frac{s\eps^2}{4}}+{\rm finite}+{\cal O}(s \eps^2)
\ee
The coefficient of the logarithm  is
uniquely determined from the obvious requirements described above.
All other pieces in the integral,
in particular, the finite part depend
completely on the chosen  function $f^\eps(t)$.

This also holds for
quadratically divergent integrals, for instance, $I_{3}^{\eps}(s)$.
On dimensional grounds,
\be
\label{dim}
I_{3}^{\eps}(s)=\frac{a}{\eps^2} + b s \log{s\eps^2}+c
s + {\cal O}(s\eps^2)
\ee
If we now derive $I_3^{\eps}$ with respect to $s$ we obtain
\be
{d\over {ds}}I_3^{\eps}(s)=b\log{s\eps^2}+{\rm finite}+
{\cal O}(\eps^2)
\ee
Which is again logarithmically divergent and, hence, we
just follow the same steps as in the previous case, finding
\be
I_{3}^{\eps}(s)=\frac{a}{\eps^2}+\frac{s}{12}\log{s\eps^2}+cs+
{\cal O}(\eps^2)
\ee

\section{Some Regulators}

We propose here some regulated propagators fulfiling the properties
(\ref{prop}) and check that indeed all of them lead to the same value for
the $F^A(s,t,u)$ amplitude.

\bigskip
\underline{Regulator 1.}
\be\Delta(x)=\frac{1}{x^2+\eps^2}\ee
This corresponds
to a $f^{\eps}(t)=e^{-t\eps^2}$. Substituting into (\ref{integ1})
gives the following values for the integrals,
\bea
&&I_{1}^{\eps}(s)=-(\log{\frac{s\eps^2}{4}}+2\gamma_E)
\nn\\
&&I_{2}^{\eps}(s)=\frac{1}{\eps^2}+\frac{s}{4}(\log{\frac{s\eps^2}{4}}+
2\gamma_E-1) \nn\\
&&I_{3}^{\eps}(s)=\frac{1}{3\eps^2}+\frac{s}{12}(\log{\frac{s\eps^2}{4}}
+2\gamma_E-1)
\eea
Putting this into (\ref{expresf}) one gets for $F^{A}(s,t,u)$ in
Minkowski space,
\bea
& F^{A}(s,t,u)=& -\frac{1}{96\pi^2 f^4}\{3s^{2}(\log{-\frac{s\eps^2
e^{2\gamma_E}
}{4}}+\frac{1}{3})+ \nn \\
& & t(t-u)(\log{-\frac{t\eps^2
e^{2\gamma_E}}{4}})+ u(u-t)(\log{-\frac{u\eps^2 e^{2\gamma_E}}{4}})\}
\eea
implying that $\beta_{1}=-2\gamma_E-\frac{1}{3}+\log{4}$ and
$\beta_{2}=
-2\gamma_E+\log{4}$. So, we recover the correct value of $\beta$ which
is $\frac{1}{3}$. On the other hand, if using this regulator, we try
to calculate the $F^{B}(s,t,u)$ part of the amplitude, which, as we have
shown, chiral symmetry requires to be zero, we find quartic and quadratic
divergences as well as finite parts. This is not a good
regulator
for the whole amplitude, but it is good enough for the part that contains
the scheme independent information, which we have identified.

\bigskip
\underline{Regulator 2.}
\be
\Delta(x)={1\over x^2}(1-
\frac{x}{a\eps} K_{1}(\frac{x}{a\eps}))\ee
has a nice expression
in momentum space:
\be \Delta(p)=\frac{1}{p^2+a\eps^2 p^4} \ee
Power counting in momentum space shows that this propagator regulates all
but the integrals that are quartically divergent. This does not influence
$F^A$ in any case and it also reproduces the value $\beta=1/3$. This
regulator can be generalized easily to an arbitrary polynomial in
momenta, that corresponds in position space to a series of modified
Bessel functions.  In particular for polynomials of order six or
greater it will be able to regulate the quartic divergences too.

\bigskip
\underline{Regulator 3.}

We can rederive the results of
dimensional
regularization simply by considering the integrals appearing in (\ref{fa}
and \ref{caixes}) in $n$ dimensions and introducing
$f^{\eps}(t)=(\frac{t}{\pi})^{n/2-2}$, with $n=4-\eps$, whose Laplace
transform
(\ref{general}) reproduces the $n$-dimensional $x$-space propagator
\be
\tg(x)= {1\over {4\pi^{n/2} x^{n-2}}} \Gamma ({n\over 2} -1)
\ee
The integrals (\ref{integ1}) in their $n$-dimensional version
reduce then to $\Gamma$-function type
integrals, whose evaluation gives the familiar result $\beta_1=11/6$,
$\beta_2=13/6$, and $\beta=1/3$\cite{Velt}.

\bigskip
\underline{Regulator 4.}

Dimensional regularization is based in analytically continuing the
amplitudes to a complex number of dimensions. We can try other type
of regulators based in analytic continuation too. We have, for instance,
checked that
\be \tg(x)={1\over x^2}J_{\eps}(\frac{\eps x}{\nu}) \ee
with $\nu$ being some arbitrary scale and $\eps$ some dimensionless
number yields $\beta=1/3$ upon analytic continuation to $\eps=0$.

\section{Differential Renormalization}

We have also investigated  the evaluation of $F^{A}(s,t,u)$ by using
the differential renormalization  method. This method was
introduced in \cite{Lati} and applied with success to rather
involved calculations in both massless and massive $\lambda
\phi^4$\cite{massive} and
QED\cite{anomal}. It works
directly in position space
and it offers some computational advantages, so it is worth investigating
its application to non-renormalizable non-linear theories like the
chiral model.

The method consists in writing the bare amplitude in position space
and proceed to regulate the short distance divergences that arise when two
points approach each other by expressing the products of propagators
as derivatives of less singular functions with a well defined Fourier
transform.
Then one performs the Fourier transform  by integration by parts and
disposes of the surface terms.
The basic identities we will need are
\bea
\label{id}
& &\frac{1}{x^4}=-\frac{1}{4}\Box \frac{\log{x^2 {\overline{M}^2}}}{x^2}
\nn \\
& &\frac{1}{x^6}=-\frac{1}{32}\Box\Box \frac{\log{x^2
{\overline{M}^2}}}{x^2} \eea
These identities are strictly valid expect for $x=0$, so in differential
renormalization one,
in a way, makes a minimal surgery on the original theory ---just one
point.

We will also need the Fourier transform
\be
\int d^4x e^{iPx} \frac{1}{x^2} \log{x^2}
{\overline{M}^2}=-\frac{4\pi^2}{p^2} \log{ \frac{p^2}{M^2}}
\ee
where ${M}={2\overline{M}}/{\gamma_E}$. $M$ is an integration constant
of the differential equations (\ref{id}). Note that the method yields
renormalized amplitudes directly; there's really no cut-off anywhere.
(In a sense, differential renormalization provides an implementation
of the BPHZ
procedure.) In fact there is really  no reason why the integration constants
for the two differential equations implied by (\ref{id}) should
be the same. In fact, generically, they are not. It is known that
in QED there are well determined relations between different scales, which
are dictated by the Ward identities of the theory. This is crucial to
recover the correct value for the axial anomaly \cite{anomal}.

By using these techniques in the evaluation of the $F^{A}(s,t,u)$
part of the amplitude one ends up, in euclidean space, with
\bea
& F^{A}(s,t,u)=& -\frac{1}{96\pi^2
f^4_{\pi}}\{3s^2\log{\frac{s}{{M_{1}}^2}}+
2t^2\log{\frac{t}{{M_{1}}^2}}+
st\log{\frac{t}{{M_{2}}^2}}+
\nn\\
& & 2u^2\log{\frac{u}{{M_{1}}^2}}+
su\log{\frac{u}{{M_{2}}^2}} \}
\eea
where one can indeed recognize the correct coefficients of the logarithms.
In deriving the previous expression we have used a different scale $M$
for each identity (\ref{id}). If one defines an adimensional
quantities $\lambda_{1}$ as the ratio of the square of
these $M$'s
\be
\lambda_{1}=\frac{{M_{1}}^2}{{M_{2}}^2}
\ee
The $\beta$ parameter obtained from the $F^{A}(s,t,u)$ amplitude (which we
know should be equal to $1/3$) is
\be
\beta=-\frac{1}{3}\log{\lambda_{1}}
\ee
Unlike QED, however, there is no way of fixing the value of $\lambda_1$
 from symmetry principles\cite{anomal}.
The standard Ward identities for the effective action gives no
information
at all to fix this $\log{\lambda_{1}}$ since there is only one diagram
at this order in chiral perturbation theory, apart
 from tadpole diagrams that
are set to zero automatically by chiral arguments.

One could try to see whether we can extract some information on
$\lambda_1$ from requiring the conservation of the chiral current,
namely $${\partial \over \partial x_\mu}\langle 0| T
j_{\mu}^{b}(x) \pi^{\alpha}(z_{1}) \pi^{\beta}(z_{2})
 \pi^{\gamma}(z_{3}) |0\rangle=
\langle 0 |T \delta^{b}{\cal{L}} \pi^{\alpha}(z_{1}) \pi^{\beta}(z_{2})
\pi^{\gamma}(z_{3}) |0\rangle$$
$$+\delta{(x-z_{1})}\langle 0 |T \delta^{b}
\pi^{\alpha}(z_{1})\pi^{\beta}(z_{2})\pi^{\gamma}(z_{3}) |0 \rangle+
\delta{(x-z_{2})}\langle 0 |T \delta^{b}
\pi^{\beta}(z_{2})\pi^{\alpha}(z_{1})\pi^{\gamma}(z_{3}) |0 \rangle$$

\be
\label{Ward}
+\delta{(x-z_{3})}\langle 0 |T \delta^{b}
\pi^{\gamma}(z_{3})\pi^{\alpha}(z_{1})\pi^{\beta}(z_{2}) |0 \rangle
\ee
where $\delta \pi^{\alpha}$ is (\ref{chiral}) and the lagrangian
is invariant under this transformation, i.e. $\delta^b {\cal L} =0$. The
associated Noether current is (expanding in powers of $1/f_\pi$)
$j_{\mu}=j_{\mu}^{(2)}+j_{\mu}^{(4)}+ ...$ with
\bea
\label{current}
&&j_{\mu}^{(2)}=\frac{f_\pi \omega^{i}}{2} \partial_{\mu} \pi^{i}
\nn\\
&&j_\mu^{(4)}=\frac{1}{3f_\pi}\omega^{i}\partial_{\mu}\pi^{k}\pi^{j}\pi^{l}(
\delta^{il}\delta^{jk}-\delta^{jl}\delta^{ik})
\eea
etc.
The first term on the r.h.s of the equality (\ref{Ward}) is zero
because of chiral symmetry invariance. To the order we are working the
matrix element of $j_\mu$ in the three pion state is the relevant one.
There are two contributions of the same order in $1/f_\pi$ to this matrix
element, namely $j_\mu^{(2)}$ with
two insertions of the interaction lagrangian, and $j_\mu^{(4)}$ with just
one insertion. The corresponding diagrams are depicted in fig. 5.

The evaluation of these expressions is tedious but quite
straigthforward. There appear integrals containing
all the different scales. At the end however
the coefficient multiplying $\log \lambda_1$ vanishes, making it
impossible to determine $\beta$ from chiral invariance arguments.
It seems that if
one uses the differential renormalization procedure
to calculate processes in chiral perturbation theory
there is a loss of information in throwing
away the surface terms that appear
when pulling out the laplacians. These surface terms
cannot be recovered, it seems, by using Ward identities. It is
perhaps worth pointing out that differential renormalization
automatically delivers renormalized amplitudes and here we are
dealing with a non-renormalizable
theory.

\section{Constructing Finite Observables}

There is only a combination of ${\cal O}(p^4)$ operators where the
logarithmic divergences cancel, namely $L_1-{1\over 2}L_2$. Therefore,
by expanding the operators of the ${\cal L}^{(4)}$ lagrangian
(\ref{lagr}) to higher
orders in $1/f_{\pi}$ one can find the observables that are finite at the
one loop level in chiral perturbation theory and, hence, their finite
parts are unambiguously predicted. It is not very complicated to
construct a computer code to perform this analysis and we have done this.
We thus have a systematic way of finding finite amplitudes
at the one loop level.
As an example we will show the first two cases.

(a) For $m=4$ external legs there are three kinematical invariants
that we denote here by
$s_{1},s_{2}$, and $s_{3}$ (the $s,t,u$ Mandlestam
variables). The only finite amplitude is proportional to the
following combination of momenta
\be
{\cal A}\sim
(\delta^{a_{1}
a_{2}}\delta^{a_{3} a_{4}} tu+\delta^{a_{1} a_{3}}\delta^{a_{2} a_{4}}
su+ \delta^{a_{1} a_{4}}\delta^{a_{2} a_{3}} st)
\ee
where one recognizes the $T(1)$ amplitude discussed at length in section
3.

(b) The case $m=6$ is a bit longer. There are 10 kinematical
invariants invariant $s_{i}$,  $i=1..10$ given by
\bea
&&s_{1}=(p_{1}+p_{2})^2\hspace{2.5mm} s_{2}=(p_{1}+p_{3})^2
\hspace{2.5mm} s_{3}=(p_{1}+p_{4})^2 \hspace{2.5mm} \nn\\
&&s_{4}=(p_{1}+p_{5})^2 \hspace{2.5mm} s_{5}=(p_{2}+p_{3})^2
\hspace{2.5mm}
s_{6}=(p_{2}+p_{4})^2\nn\\
&&s_{7}=(p_{2}+p_{5})^2 \hspace{2.5mm}
s_{8}=(p_{3}+p_{4})^2 \hspace{2.5mm} s_{9}=(p_{3}+p_{5})^2
\hspace{2.5mm} s_{10}=(p_{4}+p_{5})^2
\eea
and the (only) finite amplitude is proportional to
\bea
&&{\cal A}\sim
\delta^{a_{1}a_{2}}\delta^{a_{3}a_{4}}\delta^{a_{5}a_{6}}(s_{2}s_{3}+
s_{2}s_{5}+s_{3}s_{6}+s_{4}s_{9}+s_{4}s_{10}+s_{5}s_{6}+\nn\\
&&s_{7}s_{9}+ s_{7}s_{10})
-2(s_{2}s_{7}+
s_{2}s_{10}+s_{3}s_{7}+s_{3}s_{9}+s_{4}s_{5}+s_{4}s_{6}+ s_{5}s_{10}+
s_{6}s_{9})+
\nn
\\
&&3(s_{1}s_{9}+s_{1}s_{10}+s_{4}s_{8}+s_{7}s_{8})+4(s_{3}s_{5}+s_{2}s_{6}
-s_{1}s_{8}-s_{4}s_{7}-s_{9}s_{10}))\nn\\
&&+14\mbox{  permutations}
\eea
The permutations consist in exchanging the indices $a_{i}$ in all
possible ways and change the $s_{j}$ invariants accordingly
under an interchange of the corresponding
$p_{i}$. For instance, if one permuts $a_{4} \lra a_{5}$ it
implies to change $p_{4} \lra p_{5}$ and as a consequence
$s_{3} \lra s_{4}, s_{6} \lra s_{7}, s_{8} \lra s_{9}$. The
${\cal O}(p^4)$ lagrangian contributes to this finite amplitude
with a proportionality factor equal to $(8/3)(L_1-\frac{1}{2} L_2)$.

\clearpage

{\bf Figure Captions}
\vspace{20mm}

Fig. 1. Diagrams contributing to $\pi\pi$ scattering amplitude at tree
and one-loop order. (a): tree level from ${\cal L}^{2}$.   (b), (c) and
(d): one-loop level (s,t,u channels) from ${\cal L}^{2}$ (e): tree level
 from ${\cal L}^{4}$.
\bigskip
\bigskip

Fig. 2. Ward Identity. The black box represents the insertion of
$\delta S$ to the appropriate order. (a): tree level
diagram. (b): tadpole. (c): one-loop with one insertion of $\delta S$
and one vertex from ${\cal L}^{2}$.
\bigskip
\bigskip

Fig. 3. General one-loop diagram with {\it n} vertices and {\it m}
external legs.
\bigskip
\bigskip

Fig. 4. One of the diagrams contributing to a 6 pion process. It has
{\it n=}3 vertices and {\it m=}6 external legs.
\bigskip
\bigskip

Fig. 5. Diagrams that correspond to the Ward Identity
(\ref{Ward}). (a) and (b) contribute to the l.h.s of the W.I. and (c)
to the r.h.s. Here the black box is the insertion of the current
$j_{\mu}$ defined in eq. \ref{current} to the appropriate order.

\end{document}